\documentstyle[12pt,preprint,aps]{revtex}
\def\lord{$ \raisebox{-.3ex}{$\stackrel{<}{_{\sim}}$} $}

\preprint{NUC-MINN-00/14-T}

\title{Relativistic Viscous Fluid Description of
Microscopic Black Hole Wind}

\author{J. I. Kapusta\footnote{kapusta@physics.spa.umn.edu}}
\address{
School of Physics and Astronomy\\
University of Minnesota\\
Minneapolis, MN 55455}

\date{\today}

\begin{document}

\maketitle
\begin{abstract}

Microscopic black holes explode with their temperature varying inversely as
their mass.  Such explosions would lead to the highest temperatures in the
present universe, all the way to the Planck energy. Whether or not a
quasi-stationary shell of interacting matter undergoing radial hydrodynamic 
expansion surrounds such black holes is controversial.  In this paper 
relativistic viscous fluid equations are applied to the problem assuming 
sufficient particle interaction.  It is shown that a self-consistent picture 
emerges of a fluid just marginally kept in local thermal equilibrium; viscosity 
is a crucial element of the dynamics.

\end{abstract}

\vspace*{0.5in}

\newpage

Hawking radiation from black holes \cite{Hawk} is of fundamental interest
because it relies on the application of relativistic quantum field theory in the
presence of the strong field limit of gravity, a so far unique situation.  It is
also of great interest because of the temperatures involved.  A black hole with
mass $M$ radiates thermally with a temperature
\begin{equation}
T_h = \frac{m_{\rm P}^2}{8\pi M}
\end{equation}
where $m_{\rm P} = G^{-1/2} = 1.22\times 10^{19}$ GeV is the Planck mass.
(Units are $\hbar = c = k_{\rm B} = 1$.)  In order for the black hole
to evaporate it must have a temperature greater than that of the present-day
black-body radiation of the universe of 2.7 K = 2.3$\times 10^{-4}$ eV.
This implies that $M$ must be less than $1\%$ of the mass of the Earth,
hence the black hole most likely would have been formed primordially and not
from stellar collapse.  The black hole temperature eventually goes to infinity
as its mass goes to zero, although once $T_h$ becomes comparable to the
Planck mass the semi-classical calculation breaks down and the regime of full
quantum gravity is entered.  Only in two other situations are such enormous
temperatures achievable: in the early universe ($T$ similarly asymptotically
high) and in central collisions of heavy nuclei like gold or lead ($T = 500$ MeV
is expected at the RHIC (Relativistic Heavy Ion Collider) just completed at
Brookhaven National Laboratory and $T = 1$ GeV is expected at the LHC (Large
Hadron Collider) at CERN to be completed in 2005).  The spontaneously broken
chiral symmetry of QCD gets restored in a phase transition/rapid crossover at a
temperature around 160 MeV, while the spontaneously broken gauge symmetry in the
electroweak sector of the standard model gets restored in a phase
transition/rapid crossover at a temperature around 100 GeV.  The fact that
temperatures of the latter order of magnitude will never be achieved in a
terrestrial experiment motivates me here to study the fate of primordial black
holes during the final minutes of their lives when their temperatures have risen
to 100 GeV and above.  The fact that primordial black holes have not yet been
observed \cite{review} does not deter me in the least.

There has been some controversy over whether the particles scatter from each
other after being emitted, perhaps even enough to allow a fluid description of
the wind coming from the black hole.  First the situation will be reviewed, and
then it will be shown that a self-consistent description is afforded by the
application of relativistic viscous fluid equations, at least towards the end of
the black hole's existence and under certain assumptions.

When $T_h \ll m_{\rm e}$ (electron mass) only photons, gravitons,
and neutrinos will be created with any significant probability.  These
particles will not interact with each other but will be emitted into
the surrounding space with the speed of light.  Even when $T_h
\approx m_{\rm e}$ the Thomson cross section is too small to allow the
photons to scatter very frequently in the rarified electron-positron
plasma around the black hole.  This may change when
$T_h \approx 100$ MeV when muons and charged pions are created in
abundance.  At somewhat higher temperatures hadrons are copiously produced and
local thermal equilibrium may be achieved, although exactly how is an unsettled
issue.  Are hadrons emitted directly by the black hole?  If so, they will be
quite abundant at temperatures of order 150 MeV because their mass spectrum
rises exponentially (Hagedorn growth as seen in the Particle Data Tables
\cite{PDG}).  Because they are so massive they move nonrelativistically and may
form a very dense equilibrated gas around the black hole.  But hadrons are
composites of quarks and gluons, so perhaps quarks and gluon jets are emitted
instead?  These jets must decay into the observable hadrons on a typical length
scale of 1 fm and a typical time scale of 1 fm/c.  Once the hadrons appear they
may form an equilibrated gas around the black hole just as if they had been
produced directly albeit with some time delay.  One can find arguments both for
\cite{for} and against \cite{again} thermal equilibrium among the
strongly interacting hadrons outside the Schwarzschild radius.  Recently a
numerical study of photosphere formation around a primordial black based on a
relaxation time approximation to the Boltzmann equation has been performed for
QED and QCD interactions \cite{cline}.  It was found that significant particle
scattering would lead to a photosphere though not perfect fluid flow.  Certainly
this is a very difficult and open problem in quantum statistical mechanics, just
as it is in high energy heavy ion collisions \cite{qm}.

Let us assume that a primordial black hole is surrounded by a shell of
expanding interacting matter in approximately local thermal
equilibrium when $T_h$ is large enough.  A detailed description
of how this situation comes to be is a difficult
problem as discussed above and is not addressed in this paper.  The relativistic
imperfect fluid equations describing a steady-state, spherically symmetric flow
with no net baryon number or electric charge and neglecting gravity
(see below) are $T^{\mu\nu}_{\;\;\;\;;\nu} =$ {\em black hole source}.  The
nonvanishing components of the energy-momentum tensor in radial coordinates are
\cite{MTW}
\begin{eqnarray}
T^{00}&=& \gamma^2 (P+\epsilon) -P + v^2 \Delta T_{\rm diss} \nonumber \\
T^{0r}&=& v\gamma^2 (P+\epsilon) + v \Delta T_{\rm diss} \nonumber \\
T^{rr}&=& v^2\gamma^2 (P+\epsilon) +P + \Delta T_{\rm diss}
\end{eqnarray}
representing energy density, radial energy flux, and radial momentum flux,
respectively, in the rest frame of the black hole.  Here
$v$ is the radial velocity with $\gamma$ the corresponding Lorentz factor, $u =
v\gamma$, $\epsilon$ and $P$ are the local energy density and pressure, and
\begin{equation}
\Delta T_{\rm diss} = -\frac{4}{3}\eta \gamma^2 \left( \frac{du}{dr}
-\frac{u}{r}\right) - \zeta \gamma^2 \left( \frac{du}{dr}
+\frac{2u}{r}\right) \, ,
\end{equation}
where $\eta$ is the shear viscosity and $\zeta$ is the bulk viscosity.  A
thermodynamic identity gives $Ts = P + \epsilon$ for zero chemical potentials,
where $T$ is temperature and $s$ is entropy density.  There are two independent
differential equations of motion to solve for the functions $T(r)$ and $v(r)$.

An integral form of these equations is probably more useful since it can readily
incorporate the input luminosity $L_i$ from the black hole.  The first
represents the equality of the energy flux passing through a sphere of radius r
with the luminosity of the black hole.
\begin{equation}
4\pi r^2 T^{0r} = L_i
\end{equation}
The second follows from integrating a linear combination of the differential
equations.  It represents the combined effects of the entropy from the black
hole together with the increase of entropy due to viscosity.
\begin{equation}
4\pi r^2 u s  = 4\pi \int_{r_i}^r dr' \, r'^2 \frac{1}{T}\left[
\frac{8}{9} \eta \left( \frac{du}{dr'} - \frac{u}{r'} \right)^2
+ \zeta \left( \frac{du}{dr'} + \frac{2u}{r'} \right)^2 \right]
+ \frac{L_i}{T_h}
\end{equation}
The term $L_i/T_h$ arises from equating the entropy per unit time lost by the
black hole $-d S_{\rm h}/dt$ with that flowing into the matter.  Using the area
formula for the entropy of a black hole, $S_{\rm h} = m_{\rm P}^2 \pi r_S^2 =
4\pi M^2/m_{\rm P}^2$, and identifying $-dM/dt$ with the luminosity, the entropy
input from the black hole is obtained.

The above pair of equations are to be applied beginning at some radius $r_i$
greater than the Schwarzschild radius $r_S$, that is, outside the quantum
particle production region of the black hole.
The radius $r_i$ at which the imperfect fluid equations are first applied should
be chosen to be greater than the Schwarzschild radius, otherwise the computation
of particle creation by the black hole would be invalid.  It should not be too
much greater, otherwise particle collisions would create more entropy than is
accounted for by eq. (5).  The energy and entropy flux into the fluid come
from quantum particle creation by the black hole at temperature $T_h$, and
particle production is dominated by particles with mass less than the
temperature.  Massless particles emitted from a surface at rest have an
average outward velocity of $1/\sqrt{3}$.  Therefore it is not unreasonable
to assume the initial flow velocity is $v_i = v(r_i) = 1/\sqrt{3}$.
Furthermore, considerations of smoothness and continuity suggest that
$dv/dr = 0$ at $r_i$.  In fact one should not expect the precise choice
of initial conditions to matter at large radii; see the discussion of
the scaling solutions later in this paper.
Once the functions $s(T)$, $\eta(T)$, and $\zeta(T)$ are specified, $r_i$ and
$T_i = T(r_i)$ can be determined by the integral form of the equations of
motion, eqs. (4) and (5).  Gravitational effects are of order $r_S/r$, hence
negligible for $r > (5-10)r_S$.

A black hole has a Schwarzschild radius $r_S = 2M/m_{\rm P}^2 =
1/4\pi T_h$.  Note that $\pi T_h \cdot 2r_S = 1/2$.  Roughly, the
average thermal momentum of a massless particle times the diameter of the black
hole is 1/2.  This is just a manifestation of the uncertainty principle
applied to the creation of an excitation in a confined region of space.  The
luminosity is
\begin{equation}
L = -\frac{dM}{dt} = \alpha(M) \frac{m_{\rm P}^4}{M^2} =
64 \pi^2 \alpha(T_h) T_h^2
\end{equation}
where $\alpha(M)$ is a function reflecting the species of particles available
for creation in the gravitational field of the black hole.  It is generally
sufficient to consider only those particles with mass less than $T_h$;
more massive particles are exponentially suppressed by the Boltzmann factor.
Then
\begin{equation}
\alpha = 2.011\times 10^{-8} \left[ 4200 N_0 + 2035 N_{1/2} + 835 N_1 + 95 N_2
\right] \, .
\end{equation}
Here $N_s$ is the net number of polarization degrees of freedom for all
particles with spin $s$ and with mass less than $T_h$.  The coefficients for
spin 1/2, 1 and 2 were computed in ref. \cite{Page} and for spin 0 in ref.
\cite{Sanchez}.  In the standard model $N_0 = 4$ (Higgs), $N_{1/2} = 90$ (three
generations of quarks and leptons), $N_1 = 24$ (SU(3)$\times$SU(2)$\times$U(1)
gauge theory), and $N_2 = 2$ (gravitons).  This assumes $T_h$ is greater than
the temperature for the electroweak gauge symmetry restoration \cite{perplex}.
Numerically $\alpha(T_h > 100 \,{\rm GeV}) = 4.43\times 10^{-3}$.  Starting with
a black hole of temperature $T_h$, the time it takes to evaporate/explode is
\begin{equation}
\Delta t = \frac{m_{\rm P}^2}{3 \alpha(T_h) (8\pi T_h)^3} \, .
\end{equation}
This is also the characteristic time scale for the rate of change of the
luminosity of a black hole with temperature $T_h$.  Generally
$\Delta t \gg 1/T_h$, thus justifying the quasi-stationary
assumption \cite{jh}.  For example, a black
hole with temperature 1 TeV has a Schwarzschild radius of $1.57\times 10^{-5}$
fm, a mass of $10^{10}$ g, a luminosity of $7\times 10^{27}$ erg/s, and has 464
seconds to live.  The input luminosity to the expanding fluid $L_i$ will be less
than the total luminosity $L$, because gravitons will escape without scattering,
and neutrinos could scatter only in those regions where the temperature is
greater than about 100 GeV.  The reason is neutrino cross sections above
100 GeV will be similar to the cross sections of other fermions because of
the restoration of the spontaneously broken electroweak gauge symmetry.

Determination of the equation of state as well as the two viscosities for
temperatures ranging from MeV to TeV and more is a formidable task.  Here we
shall consider two interesting limits and then a semi-realistic situation.  A
realistic, quantitative description of the relativistic black hole wind,
including the asymptotic observed particle spectra, is left for a future
publication.

First, consider the adiabatic limit (like milk) with an equation of state
$\epsilon = aT^4$, $s = (4/3)aT^3$, and $\eta = \zeta = 0$.  This is equivalent
to assuming that the mean free paths of the particles are all small compared to
the length scale over which the temperature and other thermodynamic quantities
change significantly.  A scaling solution, valid when $\gamma \gg 1$, is $T(r) =
T_0(r_0/r)$ and $\gamma(r) = \gamma_0(r/r_0)$, where $\gamma_0 T_0 = T_h$.  The
$r_0$ is any reference radius satisfying the stated criterion.

Second, consider the highly viscous isoergic limit (like honey) in the sense
that the flow velocity approaches a limiting value $v_0$ at large $r$.  This
requires a power-like equation of state $\epsilon \propto T^{\delta}$ and
viscosities $\eta \propto \zeta \propto T^{\delta/2}$.  It results in the
scaling solution $T(r) = T_0 (r_0/r)^{2/\delta}$.  This is not very realistic: a
massless gas with dimensionless coupling constants and $\delta = 4$ would
require viscosities of order $T^2$ whereas one would expect $T^3$ on dimensional
grounds.

Now consider a semi-realistic situation with $\epsilon = aT^4$, $s = (4/3)aT^3$,
$\eta = b_ST^3$, and $\zeta = b_BT^3$.  This is typical of relativistic gases 
with dimensionless coupling constants, although quantum effects will give
logarithmic corrections \cite{tau,baym}.  A scaling solution, valid at large 
radii when $\gamma \gg 1$, is $T(r) = T_0 (r_0/r)^{2/3}$ and $\gamma(r) = 
\gamma_0 (r/r_0)^{1/3}$.  The constants are related by $36aT_0r_0 =
(32b_S + 441b_B)\gamma_0$.  This $r$-dependence of $T$
and $\gamma$ is exactly what was conjectured in ref. \cite{for}.

Is the semi-realistic situation described above really possible?  Can
approximate local thermal equilibrium, if achieved, be maintained?
The requirement is that
the inverse of the local volume expansion rate $\theta = u^{\mu}_{\;\; ;\mu}$ be
comparable to or greater than the relaxation time for thermal equilibrium
\cite{MTW}.  Expressed in terms of a local volume element $V$ and proper time
$\tau$ it is $\theta = (1/V)dV/d\tau$, whereas in the rest frame of the black
hole the same quantity can be expressed as $(1/r^2)d(r^2 u)/dr$.  Explicitly
\begin{equation}
\theta = \frac{7\gamma_0}{3r_0}\left(\frac{r_0}{r}\right)^{2/3}
= \frac{7\gamma_0}{3r_0T_0} T \, .
\end{equation}
Of prime importance in achieving and maintaining local thermal equilibrium in a
relativistic plasma are multi-body processes such as $2 \rightarrow 3$ and
$3 \rightarrow 2$, etc.  This has been well-known when calculating quark-gluon
plasma formation and evolution in high energy heavy ion collisions \cite{klaus}
and has been emphasized in ref. \cite{for} in the context of black hole
evaporation.  This is a formidable task in the standard model with its 16
species of particles.  Instead we make three estimates for the requirement that
local thermal equilibrium be maintained. The first and simplest estimate is to
require that the thermal DeBroglie wavelength of a massless particle,
$1/3T$, be less than $1/\theta$.   The second estimate is to require that
the Debye screening length for each of the gauge groups in the standard model be
less than $1/\theta$. The Debye screening length is the inverse of the Debye
screening mass $m^{\rm D}_n$ where $n =1, 2, 3$ for the gauge groups U(1),
SU(2), SU(3).  Generically $m^{\rm D}_n \propto g_nT$ where $g_n$ is the gauge
coupling constant and the coefficient of proportionality is essentially the
square root of the number of charge carriers \cite{kapbook}.  For example, for
color SU(3) $m^{\rm D}_3 = g_3 \sqrt{1+N_{\rm f}/6}\,T$
where $N_{\rm f}$ is the number of light quark flavors at the temperature $T$.
The numerical values of the gauge couplings are: $g_1 = 0.344$, $g_2 = 0.637$,
and $g_3 = 1.18$ (evaluated at the scale $m_Z$) \cite{PDG}.  So within a factor
of about 2 we have $m^{\rm D} \approx T$. The third and most relevant estimate
is the mean time between two-body collisions in the standard model for
temperatures greater than the electroweak symmetry restoration temperature.
This mean time was calculated in ref. \cite{tau} in the process of
calculating the viscosity in the relaxation time approximation.  Averaged over
all particle species in the standard model one may infer from that paper an
average time of $3.7/T$.  Taking into account multi-body reactions would
decrease that by about a factor of two to four.  All three of these estimates
are consistent within a factor of 2 or 3.  The conclusion to be drawn is that
local thermal equilibrium should be achieved when
$\theta \lord T$, or $84a/(32b_S + 441b_B) \lord 1$.  This criterion places a
minimum value on the strength of the viscosities.
Once thermal equilibrium is achieved it is not lost because $\theta/T$ is
independent of $r$.  The picture that emerges is that of an imperfect fluid just
marginally kept in local equilibrium by viscous forces.

The hot shell of matter surrounding a primordial black hole provides a
theoretical testing ground rivaled only by the big bang itself.  In addition to
the questions already raised, one may contemplate baryon number violation at
high temperature and how physics beyond the standard model might be
important in the last few minutes in the life of a primordial black hole.
Finally, such black holes may contribute to the highest energy cosmic rays whose
origin is a long-standing puzzle.  Experimental discovery of exploding black
holes will be one of the great challenges in the new millennium.

\section*{Acknowledgements}

I am grateful to G. Amelino-Camelia and A. Heckler for many illuminating
discussions and to P. Ellis for comments on the manuscript.
This work was supported by the US Department of Energy under grant
DE-FG02-87ER40328.


\begin{thebibliography} {99}

\bibitem{Hawk} S. W. Hawking, Nature (London) {\bf 248}, 30 (1974); Commun.
Math. Phys. {\bf 43}, 199 (1975).

\bibitem{review} B. J. Carr and J. H. MacGibbon, Phys. Rep. {\bf 307}, 141
(1998).

\bibitem{PDG} Particle Data Group: {\it Review of Particle Physics}, Eur. Phys.
J. {\bf C3}, 1 (1998).

\bibitem{for} A. F. Heckler, Phys. Rev. D {\bf 55}, 480 (1997); Phys. Rev. Lett.
{\bf 78}, 3430 (1997).

\bibitem{again} J. Oliensis and C. T. Hill, Phys. Lett. {\bf B143}, 92 (1984);
J. H. MacGibbon and B. R. Webber, Phys. Rev. D {\bf 41}, 3052 (1990); J. H.
MacGibbon and B. J. Carr, Astrophys. J. {\bf 371}, 447 (1991); F. Halzen, E.
Zas, J. H. MacGibbon and T. C. Weekes, Nature (London) {\bf 353}, 807 (1991).

\bibitem{cline} J. Cline, M. Mostoslavsky, and G. Servant, Phys. Rev. D
{\bf 59}, 063009 (1999).

\bibitem{qm} See, for example, the proceedings of the Quark Matter series of 
international conferences, the most recent in print being: {\it Proceedings of 
Quark Matter `99}, Nucl. Phys. {\bf A661}, (1999).

\bibitem{MTW} C. W. Misner, K. S. Thorne, and J. A. Wheeler, {\em Gravitation}
(W. H. Freeman and Company, San Francisco, 1973).

\bibitem{Page} D. N. Page, Phys. Rev. D {\bf 13}, 198 (1976).

\bibitem{Sanchez} N. Sanchez, Phys. Rev. D {\bf 18}, 1030 (1978); the
coefficient was extracted from figure 5 of this paper to within about 5\%
accuracy.

\bibitem{perplex} One may usefully contemplate whether the helicity-0 component
of the W and Z bosons below the symmetry restoring temperature is to be treated
as the third component of a free massive vector boson or as a free massive
scalar particle on account of the Higgs mechanism.   If the vexing problem of
quark/gluon versus hadron production has not attracted the curiosity of the
reader this ought to.

\bibitem{jh} J. H. MacGibbon, Phys. Rev. D {\bf 44}, 376 (1991).

\bibitem{baym} G. Baym, H. Monien, C. J. Pethick, and D. G. Ravenhall,
Phys. Rev. Lett. {\bf 64}, 1867 (1990).

\bibitem{tau}  M. E. Carrington and J. I. Kapusta, Phys. Rev. D {\bf 47}, 5304
(1993).

\bibitem{klaus} K. Kinder-Geiger, Phys. Rep. {\bf 258}, 237 (1995);
S.M.H. Wong, Phys. Rev. D {\bf 54}, 2588 (1996).

\bibitem{kapbook} J. I. Kapusta, {\em Finite Temperature Field Theory}
(Cambridge University Press, Cambridge, England, 1989).

\end{thebibliography}
\end{document}